\documentclass[journal]{IEEEtran}
\usepackage{amsthm}
\usepackage{amsfonts}
\usepackage[top=0.5in, bottom=0.6in, left=0.5in, right=0.5in]{geometry}
\usepackage{amssymb}
\ifCLASSINFOpdf
  \usepackage[pdftex]{graphicx}
  \usepackage{graphicx}
  \graphicspath{{../pdf/}{../jpeg/}{../png/}}
  \DeclareGraphicsExtensions{.pdf,.jpeg,.png}
\else
  \usepackage[dvips]{graphicx}
  \graphicspath{{../eps/}}
  \DeclareGraphicsExtensions{.eps}
\fi
\ifCLASSOPTIONcompsoc
  \usepackage[caption=false,font=normalsize,labelfont=sf,textfont=sf]{subfig}
\else
\usepackage[caption=false,font=footnotesize]{subfig}
\fi
\usepackage{multirow}
\usepackage{balance}
\usepackage{multicol}
\usepackage{epstopdf}
\usepackage{verbatim} 

\usepackage[english]{babel}
\usepackage{float}
\usepackage{xcolor}
\usepackage[linesnumbered,ruled,vlined]{algorithm2e}

\usepackage[hidelinks]{hyperref}
\usepackage{cite}
\ifCLASSINFOpdf
   \usepackage[pdftex]{graphicx}
   \else
\usepackage[dvips]{graphicx}
\fi
\usepackage[cmex10]{amsmath}
\usepackage[T1]{fontenc}
\newcommand*{\affmark}[1][*]{\textsuperscript{\dag}}
\begin{document}
\title{Variational Bayes Estimation for Affine-Precoded Superimposed Pilots in Partially Connected Dual-Wideband Tera-Hertz MU-MIMO Systems\vspace{-0.5 \baselineskip}}
\author{
\normalsize{Abhisha~Garg$^{*}$, Suraj~Srivastava$^{\dagger}$, Aditya~K.~Jagannatham$^{*}$,\\ Department of Electrical Engineering, Indian Institute of Technology Kanpur, India$^*$ \\ Department of Electrical Engineering, Indian Institute of Technology Jodhpur, India$^\dagger$\\(e-mail: abhisha20@iitk.ac.in$^*$; surajsri@iitj.ac.in$^\dagger$; adityaj@iitk.ac.in$^*$)\vspace{-3\baselineskip}}}
\maketitle
\thispagestyle{empty} 
\pagestyle{empty} 
\begin{abstract}
This work conceives two affine precoding based system models, common precoding with joint channel estimation (CP-JCE) and user-specific precoding for decoupled channel estimation (USP-DCE). Considering a \textit{dual-wideband} effected partially connected architecture, we rigorously model the terahertz (THz) multiple input multiple output (MIMO) channel for each subarray corresponding to each user by incorporating the absorption, reflection, and free-space losses. Next, to address the significant bandwidth overhead associated with conventional pilot-based channel estimation, we employ superimposed pilots. Building on this, we formulate a structured sparse channel model and develop a variational Bayesian inference algorithm that jointly estimates the channel coefficients and learns the underlying sparsity structure through hyperparameter inference, thereby enabling robust and high-precision superimposed pilot-based channel estimation under severe model uncertainty. Lastly, we compare our results for both systems and provide a trade-off analysis between them.
\end{abstract}
\vspace{-1mm}
\IEEEpeerreviewmaketitle
\vspace{-5mm}
\section{Introduction}
\vspace{-1mm}
High frequency terahertz (THz) communication, spanning from $0.3-10$ THz band \cite{jornet2011channel}, is poised to revolutionize next-generation wireless communication. In addition to the traditional frequency selectivity challenge termed \textit{frequency-wideband effect}, THz systems also experience differential delays across large antenna arrays termed \textit{spatial-wideband effect}. The coexistence of both leads to the \textit{dual-wideband effect} \cite{wang2018spatial}, giving rise to the \textit{beam split effect} caused by frequency-dependent angle-of-arrival/ angle-of-departure (AoA/AoD) variations across subcarriers. Therefore, the precise knowledge of channel state information (CSI) is crucial in such systems to enable accurate channel estimation and optimization.

Conventional cellular systems allocate dedicated pilot resources, simplifying estimation but reducing data rate due to pilot overhead. In contrast, superimposed pilots (SIP) embed low-power pilots within data, allowing simultaneous data transmission and channel estimation without extra time slots \cite{shafin2020superimposed}. Jesbin \textit{et al.} \cite{jesbin2023sparse}, proposed a sparse SIP scheme for channel estimation in OTFS systems, where all bins in a frame carry data symbols while pilot symbols are sparsely superimposed over a subset of them. Jing \textit{et al.} \cite{jing2018superimposed}, proposed an iterative channel estimation technique based on Tikhonov regularization for multi-user multiple input multiple output (MU-MIMO) scenarios, aimed at improving estimation accuracy under interference-limited conditions. Moreover, the high dimensionality of antenna arrays and the limited number of multipath components in THz channels give rise to inherent sparsity, making sparse channel estimation well suited to a compressive sensing framework. Bayesian frameworks explicitly model the sparse structure by assigning a prior distribution and updating the information via a posterior distribution conditioned on the observations. For example, the Bernoulli-Gaussian distribution \cite{wu2016block} is widely used to represent sparse signals. Moreover, sparsity-promoting priors include \textit{Gaussian-gamma distribution} \cite{cevher2009learning}, which effectively captures the sparsity pattern of the underlying signal components. The next section presents the novel contributions of this work.
\vspace{-3.5mm}
\subsection{Novel contributions of the paper}
\vspace{-1mm}
To the best of our knowledge, this is the first work that features a MU affine precoded SIP technique for a partially connected THz MIMO system impacted by the dual-wideband effect. Toward this, we propose two system models viz., the common precoding for joint channel estimation (CP-JCE) and user-specific precoding for decoupled channel estimation approaches (USP-DCE). We develop a sparse channel model for both scenarios and estimate the THz channel using \textit{variational Bayesian inference}. Notably, the Bayesian framework employed does not rely solely on the prior distribution for inference; instead, it integrates information from the likelihood function during the inference process to refine the posterior distribution. Both the system models, i.e., CP-JCE and USP-DCE, lead to performance trade-offs, which are analyzed in detail.
\begin{figure}
\centering
{\includegraphics[scale=0.12]{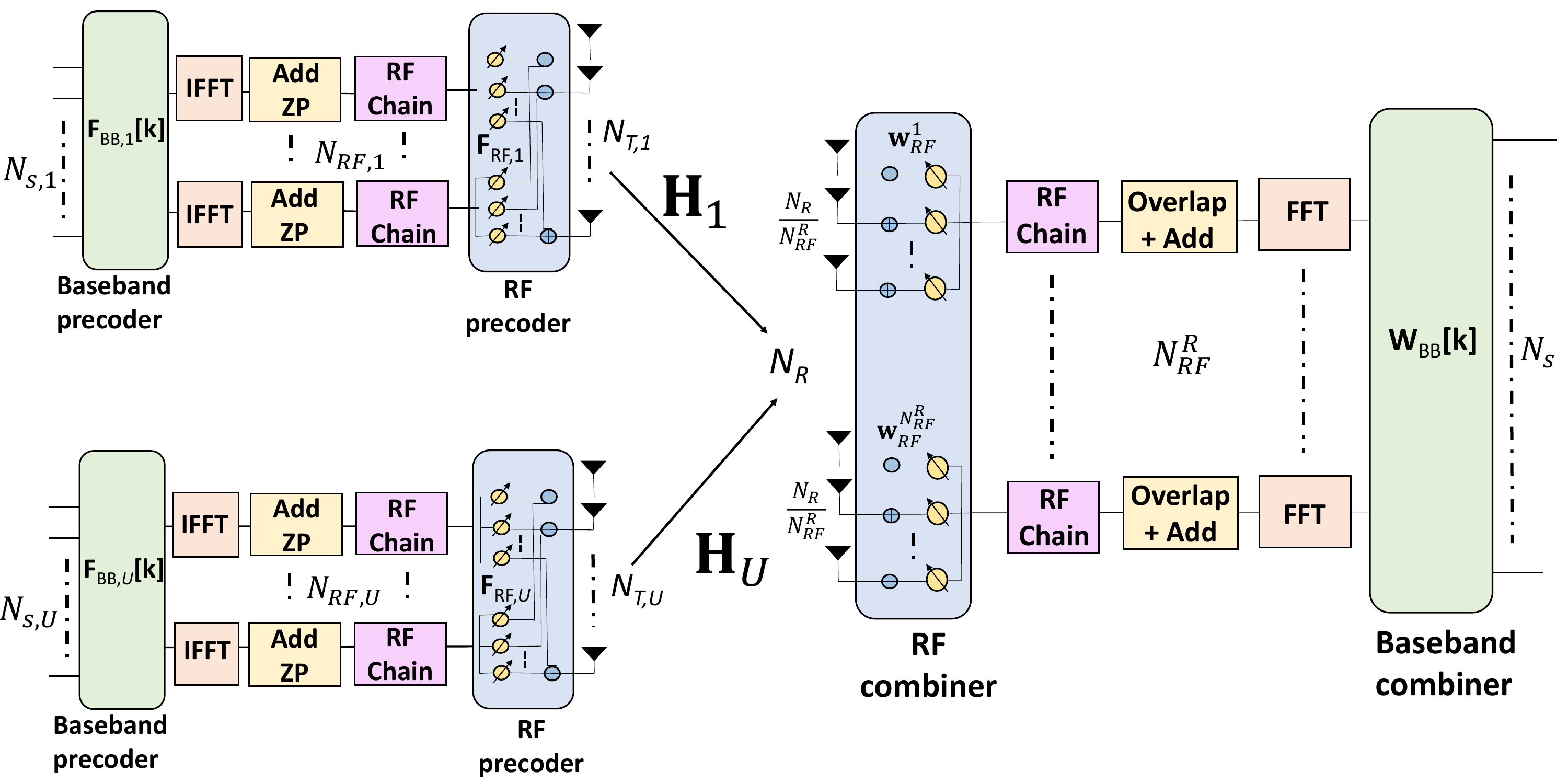}}
\vspace{-2mm}
\caption{Schematic diagram of an MU-MIMO based OFDM system.}
\label{THz_MIMO}\vspace{-1.2\baselineskip}
\end{figure}
\vspace{-2.5mm}
\section{AP-SIP based THz MU-MIMO System Model}
\vspace{-1mm}
Consider a THz MU-MIMO orthogonal frequency division multiplexing (OFDM) uplink system where a base station (BS) is equipped with $N_{sub}$ subarrays and the $l$-th subarray consists of $N_R^{l}$ antennas which serves $U$ users simultaneously with each user possessing $N_{T,u}$ transmit antennas, as shown in the Fig.\ref{THz_MIMO}. Furthermore, we have $N_{\mathrm{RF}}^R$ RF chains at the BS, while each user possess $N_{\mathrm{RF},u}$ RF chains and satisfies the relation $\sum_{u=1}^U N_{\mathrm{RF},u} \leq N_{\mathrm{RF}}^R = N_{sub} \ll N_R$. Therefore, the number of receive antennas at the BS is $N_R = \sum_{l=1}^{N_{\mathrm{RF}}^R}N_R^{l}$, while the total number of transmit antennas for all the users is $N_T = \sum_{u=1}^UN_{T,u}$ with total number of RF chains as $N_{\mathrm{RF}}^T = \sum_{u=1}^UN_{\mathrm{RF},u}$. Note that at the user equipment (UE), we have a fully-connected architecture, as shown in Fig. \ref{THz_MIMO}. At the receiver, a frequency-flat RF combiner $\mathbf{W}_{\mathrm{RF}} \in \mathbb{C}^{N_{\mathrm{RF}}^R \times N_R}$ is cascaded with a bank of frequency-selective baseband combiners $\mathbf{W}_{\mathrm{BB}}[k] \in \mathbb{C}^{N_s \times N_{\mathrm{RF}}^R}$ for each subcarrier $k$. Similarly, each user employs a frequency-flat RF precoder $\Tilde{\mathbf{F}}_{\mathrm{RF},u} \in \mathbb{C}^{N_{T,u} \times N_{\mathrm{RF},u}}$ followed by frequency-selective baseband precoders $\Tilde{\mathbf{F}}_{\mathrm{BB},u}[k] \in \mathbb{C}^{N_{\mathrm{RF},u} \times N_{s,u}}$. The RF components satisfy constant-modulus constraints $\left|\mathbf{W}_{\mathrm{RF}}(a,b)\right| = \frac{1}{\sqrt{N_R}}$, $\left|\mathbf{F}_{\mathrm{RF}_u}(a,b)\right| = \frac{1}{\sqrt{N_{T,u}}}$. 

We consider a wideband THz channel $\mathbf{H}_{d,u}^{l} \in \mathbb{C}^{N_R^{l} \times N_{T,u}}, \; 0 \leq d \leq D$, for the $d$-th tap of user $u$ on subarray $l$. Therefore, the overall channel matrix of the partially connected architecture at the BS corresponding to $u$-th user is $\mathbf{H}_u[k] = \big[\big(\mathbf{H}_u^1[k]\big)^T \: \big(\mathbf{H}_u^2[k]\big)^T \; \cdots \; \big(\mathbf{H}_u^{N_{\mathrm{RF}}^R}[k]\big)^T \big]^T \in \mathbb{C}^{N_R \times N_{T}}$, where $\mathbf{H}_u^{l}[k] \in \mathbb{C}^{N_R^{l} \times N_{T,u}}$ denotes the channel between $u$-th user and $l$-th subarray. Moreover, at the receiver, the analog combiner possess the structure $\mathbf{W}_{\mathrm{RF}} = \mathrm{blkdiag}\big(\Bar{\mathbf{w}}_{\mathrm{RF}}^{s_1} \: \Bar{\mathbf{w}}_{\mathrm{RF}}^{s_2} \: \cdots \: \Bar{\mathbf{w}}_{\mathrm{RF}}^{s_{N_{\mathrm{RF}}^R}} \big)$ where $\Bar{\mathbf{w}}_{\mathrm{RF}}^{l} \in \mathbb{C}^{N_{\mathrm{BS}}^{l} \times 1}$. Thus, the frequency domain output $\Tilde{\mathbf{Y}}_{\mathrm{MU},m}[k] \in \mathbb{C}^{N_{\mathrm{RF}}^R \times Q}$ corresponding to the $m$-th block at the $k$-th subcarrier is given as
\vspace{-2mm}
\begin{equation}
    \begin{aligned}
    \Tilde{\mathbf{Y}}_{\mathrm{MU},m}[k] = \mathbf{W}_{\mathrm{RF},m}^H \mathbf{H}_{\mathrm{MU}}[k] &\mathbf{F}_{\mathrm{RF},m} \Tilde{\mathbf{S}}_{\mathrm{MU},m}[k] + \\ & \mathbf{W}_{\mathrm{RF},m}^H \Tilde{\mathbf{V}}_{\mathrm{MU},m}[k], \label{SystemModel}
\end{aligned}
\end{equation}
where $\mathbf{H}_{\mathrm{MU}}[k] \in \mathbb{C}^{N_R \times N_{T}}$ represents the concatenated channel for all the $U$ users and is given as $\mathbf{H}_{\mathrm{MU}} = \left[\mathbf{H}_1[k] \: \mathbf{H}_2[k] \: \cdots \: \mathbf{H}_U[k]\right]$, $\mathbf{F}_{\mathrm{RF},m} \in \mathbb{C}^{N_T \times N^T_{\mathrm{RF}}}$ represents the stacked RF precoder corresponding to $U$ users for the $m$-th block and is formulated as
$\mathbf{F}_{\mathrm{RF},m} = \mathrm{blkdiag}\big(\Tilde{\mathbf{F}}_{\mathrm{RF},1,m} \: \cdots \: \Tilde{\mathbf{F}}_{\mathrm{RF},U,m}\big)$. The quantity $\Tilde{\mathbf{V}}_{\mathrm{MU},m}[k] \in \mathbb{C}^{N_{\mathrm{RF}}^R \times Q}$ represents the additive white Gaussian noise (AWGN) the elements of which are modeled as independent and identically distributed (i.i.d) complex Gaussian with $\mathcal{CN}(0,\sigma^2)$ whereas $\Tilde{\mathbf{S}}_{\mathrm{MU},m}[k] \in \mathbb{C}^{N_{\mathrm{RF}}^T \times Q}$ represents the combined affine-precoded transmit superimposed training and data symbols corresponding to all the users and given as
$\Tilde{\mathbf{S}}_{\mathrm{MU},m}[k] = \big[\Tilde{\mathbf{S}}_{1,m}[k]; \Tilde{\mathbf{S}}_{2,m}[k]; \: \cdots \: ; \Tilde{\mathbf{S}}_{U,m}[k]\big]$. Moreover, $\Tilde{\mathbf{S}}_{u,m}[k] \in \mathbb{C}^{N_{\mathrm{RF},u} \times Q}$ represents the superimposed training and data symbols corresponding to $u$-th user given by
\vspace{-2mm}
\begin{align}
    \Tilde{\mathbf{S}}_{u,m}[k] = \Tilde{\mathbf{S}}_{d,u,m}[k] \mathbf{X}_{d,m} + \Tilde{\mathbf{S}}_{p,u,m}[k]\mathbf{X}_{p,m}, \label{superimposed}
\end{align}
where $\Tilde{\mathbf{S}}_{d,u,m}[k] \in \mathbb{C}^{N_{\mathrm{RF},u} \times Q_1}$ represents the data symbols corresponding to the $u$-th user while $\Tilde{\mathbf{S}}_{p,u,m}[k] \in \mathbb{C}^{N_{\mathrm{RF},u} \times Q_2}$ represents the pilot symbols. Note that, $Q$ represents the superimposed block length, while $Q_1$ represents the block length corresponding to the data symbols and $Q_2$ represents the block length corresponding to pilot symbols, obeying $Q = Q_1 + Q_2$. The quantities $\mathbf{X}_{d,m} \in \mathbb{C}^{Q_1 \times Q}$, $\mathbf{X}_{p,m} \in \mathbb{C}^{Q_2 \times Q}$ represent the precoding matrices corresponding to data and training symbols, respectively, which can be readily obtained through unitary matrices as
\vspace{-2mm}
\begin{align}
    &\mathbf{X}_{d,m} \mathbf{X}_{d,m}^H = \mathbf{I}_{Q_1 \times Q_1} \quad\quad \mathbf{X}_{p,m} \mathbf{X}_{p,m}^H = \mathbf{I}_{Q_2 \times Q_2} \notag\\
    &\mathbf{X}_{d,m} \mathbf{X}_{p,m}^H = \mathbf{0}_{Q_1 \times Q_2} \quad\quad \mathbf{X}_{p,m} \mathbf{X}_{d,m}^H = \mathbf{0}_{Q_2 \times Q_1}. \label{precoding_matrix}    
\end{align}
However, it is interesting to note that these unitary matrices, $\mathbf{X}_{d,m}$ and $\mathbf{X}_{p,m}$, possess a special structure that can facilitate easier decoupling across users.
\vspace{-3mm}
\subsection{Common precoding with joint channel estimation (CP-JCE)}
\vspace{-1mm}
For this scenario, we set $\mathbf{X}_{p,u,m} = \mathbf{X}_{p,m}$ and $\mathbf{X}_{d,u,m} = \mathbf{X}_{d,m}, \: \forall \: u$. In order to obtain the pilot model, we post-multiply the received output $\Tilde{\mathbf{Y}}_{\mathrm{MU},m}[k]$ by the common training precoding matrix $\mathbf{X}_{p,m}^H$. Substituting Eq. \eqref{superimposed} into Eq. \eqref{SystemModel}, we get
\vspace{-1mm}
\begin{align}
    \Tilde{\mathbf{Y}}_{\mathrm{MU},m}[k]\mathbf{X}_{p,m}^H = &\mathbf{W}_{\mathrm{RF},m}^H \mathbf{H}_{\mathrm{MU}}[k] \mathbf{F}_{\mathrm{RF},m} \Tilde{\mathbf{S}}_{\mathrm{MU}_{\mathcal{P}}, m}[k] + \notag \\ &\quad\quad\quad\quad\mathbf{W}_{\mathrm{RF},m}^H \Tilde{\mathbf{V}}_{\mathrm{MU},m}[k]\mathbf{X}_{p,m}^H, \label{SamePrecoding}
\end{align}
where $\Tilde{\mathbf{S}}_{\mathrm{MU}_\mathcal{P},m}[k] \in \mathbb{C}^{N_{\mathrm{RF}}^T \times Q_2}$ represents the concatenated pilot matrix for all the $U$ users and is given as $\Tilde{\mathbf{S}}_{\mathrm{MU}_\mathcal{P},m}[k] = \big[\Tilde{\mathbf{S}}_{p,1,m}[k]; \: \Tilde{\mathbf{S}}_{p,2,m}[k]; \: \cdots \: ;\Tilde{\mathbf{S}}_{p,U,m}[k]  \big]$. Let $\mathbf{V}_{\mathrm{MU}_\mathcal{P},m}[k] = \mathbf{W}_{\mathrm{RF},m}^H \Tilde{\mathbf{V}}_{\mathrm{MU},m}[k]\mathbf{X}_{p,m}^H \in \mathbb{C}^{N_{\mathrm{RF}}^R\times Q_2}$ represent the equivalent noise matrix, $ \mathbf{Y}_{\mathrm{MU}_\mathcal{P},m}[k] = \Tilde{\mathbf{Y}}_{\mathrm{MU},m}[k]\mathbf{X}_{p,m}^H \in \mathbb{C}^{N_{\mathrm{RF}}^R \times Q_2}$ represent the equivalent received signal and $\mathbf{S}_{\mathrm{MU}_\mathcal{P},m}[k] = \mathbf{F}_{\mathrm{RF},m} \Tilde{\mathbf{S}}_{\mathrm{MU}_\mathcal{P},m}[k] \in \mathbb{C}^{UN_{T,u} \times Q_2}$ represent the equivalent pilot symbols. The noise covariance matrix $\mathbb{E}\big\{\mathbf{V}_{\mathrm{MU}_\mathcal{P},m}[k]\mathbf{V}_{\mathrm{MU}_\mathcal{P},m}^H[k]\big\} = \mathcal{CN}\big(\mathbf{0}_{N_{\mathrm{RF}}^R \times 1}, \mathbf{C}_m\big)$ with $\mathbf{C}_m \in \mathbb{C}^{N_{\mathrm{RF}}^R \times N_{\mathrm{RF}}^R}$.
\vspace{-3mm}
\subsection{User-specific precoding with decoupled channel estimation (USP-DCE)}
We enforce another condition in addition to \eqref{precoding_matrix} which ensures the orthogonality of pilot and data precoding matrices across users, i.e., $\mathbf{X}_{p,u_1,m} \mathbf{X}_{p,u_2,m}^H = \mathbf{0}$ and $\mathbf{X}_{d,u_1,m} \mathbf{X}_{d,u_2,m}^H = \mathbf{0}, \forall \, u_1 \neq u_2$. Substituting this condition into Eq. \eqref{SystemModel} and post-multiplying by the pilot precoding matrix $\mathbf{X}_{p,u,m}$ corresponding to the $u$-th user, we obtain
\begin{align}
    \Tilde{\mathbf{Y}}_{\mathrm{MU},m}[k]\mathbf{P}_{p,m}^H = &\mathbf{W}_{\mathrm{RF},m}^H \mathbf{H}_u[k] \mathbf{F}_{\mathrm{RF},u,m} \Tilde{\mathbf{S}}_{p,u,m}[k] + \notag \\ &\quad\quad\quad\quad\mathbf{W}_{\mathrm{RF},m}^H \Tilde{\mathbf{V}}_{\mathrm{MU},m}[k]\mathbf{P}_{p,m}^H, \label{DiffPrecoding}
\end{align}
where $\mathbf{Y}_{p,u,m}[k] = \Tilde{\mathbf{Y}}_{\mathrm{MU},m}[k]\mathbf{P}_{p,m}^H \in \mathbb{C}^{N_{\mathrm{RF}}^R \times Q_2}$ represents the received signal corresponding to the $u$-th user while $\mathbf{V}_{p,u,m} = \mathbf{W}_{\mathrm{RF},m}^H \Tilde{\mathbf{V}}_{\mathrm{MU},m}[k]\mathbf{P}_{p,m}^H \in \mathbb{C}^{N_{\mathrm{RF}}^R \times Q_2}$ represents the effective noise. Additionally, the noise covariance matrix $\mathbb{E}\left\{\mathbf{V}_{p,u,m}[k]\mathbf{V}_{p,u,m}^H[k]\right\} = \mathcal{CN}\big(\mathbf{0}_{N_{\mathrm{RF}}^R \times 1}, \mathbf{C}_{u,m}\big)$ with $\mathbf{C}_{u,m} \in \mathbb{C}^{N_{\mathrm{RF}}^R \times N_{\mathrm{RF}}^R}$. Let $\left[\mathbf{A}_{p,m}, \cdots, \mathbf{P}_{p,m} \right]$ represents the precoding matrix corresponding to each user which are generated from a large discrete Fourier transform (DFT) matrix with orthogonal columns. These conditions can be practically satisfied by selecting the precoding matrices as sub-matrices of an unitary matrix, where each semiunitary submatrix corresponds to a specific user. Furthermore, it can be observed that the model in Eq. \eqref{DiffPrecoding} decouples the received signals for individual users.
\vspace{-2mm}
\section{MU-THz Channel Model and Sparse Model}
\vspace{-1mm}
The array response vector $\Tilde{\mathbf{a}}^{l}_R(\theta,f_k)$ considering the spatial-wideband effect \cite{wang2018spatial} for the $l$-th subarray with $N_R^{l}$ elements, can be expressed as
\vspace{-2mm}
\begin{align}
    {\mathbf{a}}^{l}(\theta,f_k) = \frac{1}{\sqrt{N_R^{l}}}\big[1 \: e^{-j \pi \delta_k \cos{\theta}} \: \cdots \: e^{-j(N_R^{l}-1) \pi \delta_k \cos{\theta}} \big]^T, \notag
\end{align}
where $\delta_k = \frac{f_k}{f_c}$ represents the relative subcarrier frequency with $f_k \triangleq \left(k-\frac{K+1}{2}\right)\frac{B}{K}$. Furthermore, the quantity $\theta^{l}$ represents the incident angle, while $B$ denotes the bandwidth. Additionally, the THz channel includes a line-of-sight (LoS) component and few non-LoS (NLoS) components. Therefore, the channel model for subcarrier $k$ at the $l$-th subarray can be expressed as $\mathbf{H}_u^{l}[k] = \mathbf{H}_u^{l, \mathrm{LoS}} + \mathbf{H}_u^{l, \mathrm{NLoS}}$ given as
\begin{align}
    \mathbf{H}_u^{l,\mathrm{LoS}}[k] = \sqrt{N_{T,u}N_R^{l}}\alpha(f_k,d)\gamma_\tau\mathfrak{B}_{T,u}\mathfrak{B}_{R}^{l}  {\mathbf{a}}^{l}_R\left(\theta,f_k\right)  \mathbf{a}_u^H\left(\phi,f_k\right),
\end{align}
\vspace{-7mm}
\begin{align}
    \mathbf{H}_u^{l,\mathrm{NLoS}}[k] = \sqrt{\frac{N_{T,u}N_R^{l}}{N_{\mathrm{NLoS}}N_{\mathrm{ray}}}}\sum_{\ell = 1}^{N_{\mathrm{NLoS}}}\sum_{\j = 1}^{N_{\mathrm{ray}}}\alpha_{\ell,\j}(f_k,d_{\ell,\j})\\ \notag \gamma_{\tau_{\ell,\j}} \mathfrak{B}_{T,u}\mathfrak{B}_{R}^{l}{\mathbf{a}}^{l}_R\left(\theta_{\ell,\j},f_k\right)  \mathbf{a}_u^H\left(\phi_{\ell,\j},f_k\right),
\end{align}
and $\gamma_{\tau_{\ell,\j}} = \sum_{\jmath = 0}^{K-1}p(\jmath T_s - \tau_{\ell,\j})e^{-j\frac{2 \pi k \jmath}{K}}, \: \forall \, k,\jmath$. The quantities $\mathfrak{B}_{T,u}$ and $\mathfrak{B}_R$ represent the transmit and receive antenna gains, respectively. Furthermore, the variables $p(.), \tau_{(.)}, N_{\mathrm{NLoS}}, N_{\mathrm{ray}}$, $\alpha(f_k,d)$ denotes the pulse shaping filter, delay, number of NLoS components, number of diffused rays and complex-path gain respectively \cite{garg2025bayesian}. Additionally, the molecular absorption coefficient is calculated using HITRAN database \cite{garg2025bayesian}.

The extended virtual channel representation corresponding to the $l$-th subarray can be expressed as $\mathbf{H}_u^{l}[k] = \mathbf{A}_{R}^{l}\left(\Theta_R,f_k\right) \mathbf{H}_{b,u}^{l}[k]\mathbf{A}_{T,u}^H\left(\Phi_{T,u},f_k\right)$ where $\mathbf{A}_{R}^{l}\left(\Theta_R,f_k \right) \in \mathbb{C}^{N_R^{l} \times G_R^{l}}$, $\mathbf{A}_{T,u}\left(\Phi_{T,u},f_k\right) \in \mathbb{C}^{G_{T,u} \times N_{T,u}}$ represents the receive and transmit array manifold given as
\begin{align}
    \mathbf{A}_R^{l}(\Theta_R,f_k) = \big[{\mathbf{a}}_R^{l}({\theta}_1,f_k) \: {\mathbf{a}}_R^{l}({\theta}_2,f_k) \: \cdots \: {\mathbf{a}}_R^{l}({\theta}_{G_R},f_k)\big],
\end{align}
\vspace{-7mm}
\begin{align}
    \mathbf{A}_{T,u}(\Phi_{T,u},f_k) = \big[\mathbf{a}_{T,u}(\phi_{1,u},f_k) \:  \cdots \:  \mathbf{a}_{T,u}(\phi_{G_{T,u}},f_k) \big],
\end{align}
and $G_{T,u}$ and $G_R^{l}$ represent the transmit and receive bin with $\Phi_{T,u}$ and $\Theta_R^{l}$ as directional cosines \cite{garg2025bayesian} and $\mathbf{H}_{b,u}^{l}$ represents the sparse channel corresponding to $l$-th subarray which needs to be estimated. Furthermore, the overall extended virtual channel corresponding to each user for all the subarrays can be given as $\mathbf{H}_u[k] = \mathbf{A}_R\left(\Theta_R,f_k \right)\mathbf{H}_{b,u}[k]\mathbf{A}_{T,u}^H\left(\Phi_{T,u},f_k\right)$ where $\mathbf{A}_R\left(\Theta_R,f_k \right) = \mathrm{blkdiag}\big(\mathbf{A}_R^{1}\left(\Theta_R,f_k \right) \: \mathbf{A}_R^{2}\left(\Theta_R,f_k \right) \: \cdots \: \mathbf{A}_R^{{N_{RF}^R}}\left(\Theta_R,f_k \right)\big)$ and $\mathbf{H}_{b,u}[k] = \big[\mathbf{H}_{b,u}^{1}[k];\mathbf{H}_{b,u}^{2}[k];\cdots;\mathbf{H}_{b,u}^{{N_{RF}^R}}[k]\big]$. Let $\Tilde{\mathbf{\Psi}}_u[k]\in \mathbb{C}^{N_RN_{T,u} \times G_R G_{T,u}}$ represent the \textit{sparsifying dictionary} which is defined as $\mathbf{A}_{T,u}^*\left(\Phi_{T,u},f_k\right) \otimes \mathbf{A}_R\left(\Theta_R,f_k\right)$. The vectorized channel frequency response (CFR) for all the $U$ users can be determined as $\mathbf{h}_{\mathrm{MU}}[k] = \mathrm{blkdiag}\underbrace{\big(\Tilde{\mathbf{\Psi}}_1[k] \: \Tilde{\mathbf{\Psi}}_2[k] \: \cdots\: \Tilde{\mathbf{\Psi}}_U[k]\big)}_{\mathbf{\Psi}_{\mathrm{MU}}[k]}\underbrace{\big[\mathbf{h}_{b,1}^T[k] \mathbf{h}_{b,2}^T[k]  \cdots \mathbf{h}_{b,U}^T[k]\big]^T}_{\mathbf{h}_{b,\mathrm{MU}}[k]},$
where $\mathbf{h}_{\mathrm{MU}}[k] = \mathrm{vec}(\mathbf{H}_{\mathrm{MU}}[k])$, $\mathbf{\Psi}_{\mathrm{MU}}[k] \in \mathbb{C}^{N_R N_{T} \times G_R G_{T}}$ represents the joint sparsifying dictionary and $\mathbf{h}_{b,\mathrm{MU}}[k] \in \mathbb{C}^{{G}_R G_{T} \times 1}$ represents the joint beamspace output for all the users where $\sum_{u=1}^U G_{T,u} = G_T$. After employing the $\mathrm{vec}(.)$ operator in \eqref{SamePrecoding}, the MU equivalent sensing matrix, vectorized received signal, vectorized noise signal can be given as  $\Tilde{\boldsymbol{\Lambda}}_{\mathrm{MU}_\mathcal{P},m}[k] = \big(\mathbf{S}_{\mathrm{MU}_\mathcal{P},m}^T[k] \otimes \mathbf{W}_{\mathrm{RF},m}^H \big) \in \mathbb{C}^{Q_2N_{\mathrm{RF}}^R \times N_{T} N_R}$, $\mathbf{y}_{\mathrm{MU}_\mathcal{P},m}[k] = \mathrm{vec}\left(\mathbf{Y}_{\mathrm{MU},m}[k]\right)$, $\mathbf{v}_{\mathrm{MU}_\mathcal{P},m}[k] = \mathrm{vec}\left(\mathbf{V}_{\mathrm{MU}_\mathcal{P},m}[k]\right)$ respectively. Therefore, the joint CP-JCE based channel model is given as
\vspace{-4mm}

\small
\begin{align}
    \mathbf{y}_{\mathrm{MU}_\mathcal{P},m}[k] = \underbrace{\Tilde{\boldsymbol{\Lambda}}_{\mathrm{MU}_\mathcal{P},m}[k]\mathbf{\Psi}_{\mathrm{MU}}[k]}_{\boldsymbol{\Lambda}_{\mathrm{MU}_{\mathcal{P}},m}[k]}\mathbf{h}_{b,\mathrm{MU}}[k]+\mathbf{v}_{\mathrm{MU}_\mathcal{P},m}[k]. \label{MU-sparse}
\end{align}
\normalsize
Furthermore, to develop a compatible model for the $k$-th subcarrier, we concatenate the outputs for all the $M$ blocks
\begin{align}
    \underbrace{\begin{bmatrix}
        \mathbf{y}_{\mathrm{MU}_\mathcal{P},1}[k] \\ \vdots \\ \mathbf{y}_{\mathrm{MU}_\mathcal{P},M}[k]
    \end{bmatrix}}_{\mathbf{y}_{\mathrm{MU}_\mathcal{P}}[k]} = \underbrace{\begin{bmatrix}
        \boldsymbol{\Lambda}_{\mathrm{MU}_\mathcal{P},1}[k] \\ \vdots \\ \boldsymbol{\Lambda}_{\mathrm{MU}_\mathcal{P},M}[k]
    \end{bmatrix}}_{\boldsymbol{\Lambda}_{\mathrm{MU}_\mathcal{P}}[k]}\mathbf{h}_{b,\mathrm{MU}} + \underbrace{\begin{bmatrix}
        \mathbf{v}_{\mathrm{MU}_\mathcal{P},1}[k] \\  \vdots \\ \mathbf{v}_{\mathrm{MU}_\mathcal{P},M}[k]
    \end{bmatrix}}_{\mathbf{v}_{\mathrm{MU}_\mathcal{P}}[k]}, \notag
\end{align}
where $\mathbf{C}_w = \mathbb{E}\{\mathbf{v}_{\mathrm{MU}_{\mathcal{P}}}[k]\mathbf{v}_{\mathrm{MU}_{\mathcal{P}}}^H[k]\}$ is the noise covariance. Following a similar procedure, one can obtain the sparse channel estimation model for the USP-DCE approach, which is given as
\vspace{-2mm}
\begin{align}
    \mathbf{y}_{p,u,m}[k] = \boldsymbol{\Lambda}_{p,u,m}[k]\mathbf{h}_u[k] + \mathbf{v}_{p,u,m}[k], \label{me-2}
\end{align}
where $\mathbf{y}_{p,u,m}[k] = \mathrm{vec}(\mathbf{Y}_{u,m}[k])$, $\mathbf{h}_u[k] = \mathrm{vec}(\mathbf{H}_u[k])$ and $\mathbf{v}_{p,u,m}[k] = \mathrm{vec}(\mathbf{V}_{u,m}[k])$. Furthermore, $\boldsymbol{\Lambda}_{p,u,m}[k] = (\Tilde{\mathbf{S}}_{p,u,m}^T[k]\mathbf{F}_{\mathrm{RF},u,m}^T \otimes \mathbf{W}_{\mathrm{RF},m}^H) \in \mathbb{C}^{Q_2 N_{\mathrm{RF}}^R \times N_{T,u}N_R}$ represents the equivalent sensing matrix corresponding to $u$-th user. Similarly, one can obtain the stacked model for the USP-DCE approach for all the $M$ blocks which is omitted here due to space constraints. Interestingly, both \eqref{MU-sparse} and \eqref{me-2} represent sparse channel estimation models for the joint channel $\mathbf{h}_{b,\mathrm{MU}}$ of all the users and $\mathbf{h}_{b,u}$ of the $u$-th user. Consequently, we provide a detailed sparse estimation approach for the CP-JCE model of \eqref{MU-sparse}. The next section introduces a variational Bayesian inference algorithm for sparse CSI estimation.
\vspace{-2mm}
\section{Variational Inference based sparse estimation}
Bayesian probabilistic models are learning models that assume the unknown variables obeying certain posterior distribution, which is composed of prior distributions reflecting our belief on the variables before any observation, and the likelihood function describing the observations. In this regard, let the parameterized Gaussian prior $g(\mathbf{h}_{b,\mathrm{MU}}[k];\boldsymbol{\Gamma}_{k,\mathrm{MU}})$ corresponding to the $k$-th subcarrier be defined as
\vspace{-4mm}

\small
\begin{align}
    g(\mathbf{h}_{b,\mathrm{MU}}[k];\boldsymbol{\Gamma}_{k,\mathrm{MU}}) = \prod_{\imath=1}^{UG_RG_{T,u}}(\pi \gamma_{k,\imath})^{-1} \mathrm{exp}\Big(-\frac{|\mathbf{h}_{b,\mathrm{MU}}[k](\imath)|^2}{\gamma_{k,\imath}} \Big), \label{para-prior}
\end{align}
\normalsize
where $\gamma_{k,\imath}$ represents the $\imath$-th hyperparameter corresponding to $k$-th subcarrier and $\boldsymbol{\Gamma}_{k,\mathrm{MU}} = \mathrm{diag}(\gamma_{k,1},\gamma_{k,2},\cdots,\gamma_{k,UG_RG_{T,u}}) \in \mathbb{R}^{UG_RG_{T,u} \times U G_R G_{T,u}}$ denotes the hyperparameter matrix. To promote sparsity, we adopted the \textit{Gaussian-Gamma} prior, which is widely used in the machine learning literature \cite{cevher2009learning}, where the Gamma prior is conjugate to the Gaussian prior.\\
\underline{\textbf{Sparsity-enhancing Prior:}}
We constrained the hyperparameters $\bar{\gamma}_k = (\gamma_{k,1},\cdots,\gamma_{k,UG_RG_{T,u}})^T$ by treating them as random variables and imposing a \textit{Gamma prior distribution} as
\vspace{-2mm}
\begin{equation}
    \begin{aligned}
     g(\bar{\gamma}_k; \kappa, \chi) &=  \mathrm{Gamma}(\gamma_{k,\imath}|\kappa_\imath,\chi_\imath), \\
     & = \frac{\chi^\kappa}{\Upsilon(\kappa)}\gamma_{k,\imath}^{(\kappa-1)}\mathrm{exp}(-\chi\gamma_{k,\imath}), \: \gamma_{k,\imath} > 0,
\end{aligned}
\end{equation}
where $\kappa > 0$ and $\chi > 0$ while $\Upsilon(\kappa) = \int_0^\infty u^{\kappa-1} e^{-u} du$. Therefore, the joint distribution over all the hyperparameters can be given as $g(\bar{\gamma}_k) = \prod_{\imath = 1}^{U G_R G_{T,u}}g(\gamma_{k,\imath}|\kappa_\imath, \chi_\imath)$.\\
\underline{\textbf{Modelling unknown noise precision:}} We assume the Gamma prior distribution $g(\zeta; {\varphi},{\varsigma}) = \mathrm{Gamma}(\zeta|\varphi,\varsigma)$
over the noise precision $\frac{1}{\sigma^2}$ to effectively adapt the actual noise condition in the observed data. Note that, the high dimensional integral $g(\mathbf{y}_{\mathrm{MU}_{\mathcal{P}}}[k]) = \int g(\mathbf{y}_{\mathrm{MU}_{\mathcal{P}}}[k],\mathcal{J}) d\mathcal{J}$ is mathematically intractable, where $\mathcal{J} = \{\mathbf{h}_{b,\mathrm{MU}}[k],\bar{\boldsymbol{\gamma}}_k,\zeta\}$ represents the set of latent variables. To address this, we employ variational Bayes inference, which seeks an approximate distribution that is close to the posterior distribution in Kullback-Leibler (KL) divergence and formulated as Eq. \eqref{KL_eqn}, where $f(\mathcal{J})$ represents a tractable distribution.
\begin{figure*}[hbt!]
    \begin{align}
    \ln g(\mathbf{y}_{\mathrm{MU}_\mathcal{P}}[k])= \int \underbrace{f(\mathcal{J}) \ln \frac{g(\mathbf{y}_{\mathrm{MU}_\mathcal{P}}[k],\mathcal{J}) d\mathcal{J}}{f(\mathcal{J})}}_{\mathcal{L}(f)} - \int \underbrace{f(\mathcal{J}) \ln \frac{g(\mathcal{J}|\mathbf{y}_{\mathrm{MU}_\mathcal{P}}[k]) d\mathcal{J}}{f(\mathcal{J})}}_{\mathrm{KL}(f(\mathcal{J})\parallel g(\mathcal{J}|\mathbf{y}_{\mathrm{MU}_\mathcal{P}}[k]))},\label{KL_eqn}
\end{align}
\hrulefill \vspace{-1.5\baselineskip}
\end{figure*}
\begin{figure*}[hbt!]
\begin{align}              
 &g(\mathbf{h}_{b,\mathrm{MU}}[k], \bar{\gamma}_k, \zeta| \mathbf{y}_{\mathrm{MU}_\mathcal{P}}[k]) = \frac{g(\mathbf{y}_{\mathrm{MU}_\mathcal{P}}[k]|\mathbf{h}_{b,\mathrm{MU}}[k],\zeta)g(\mathbf{h}_{b,\mathrm{MU}}[k]|\gamma_k)g(\bar{\gamma}_k)g(\zeta)}{g(\mathbf{y}_{\mathrm{MU}_\mathcal{P}}[k])}, \label{posterior}\\
 &g(\mathbf{h}_{b,\mathrm{MU}}[k],\bar{\gamma}_k,\zeta|\mathbf{y}_{\mathrm{MU}_\mathcal{P}}[k]; \kappa, \chi, \varphi, \varsigma) \approx \prod_{k=1}^K f(\mathbf{h}_{b,\mathrm{MU}}[k], \bar{\gamma}_k, \zeta) = \prod_{k=1}^K f(\mathbf{h}_{b,\mathrm{MU}}[k]) f(\bar{\gamma}_k) f(\zeta). \label{VBLapprox}
\end{align}
\hrulefill \vspace{-1.5\baselineskip}
\end{figure*}
\begin{figure*}[hbt!]
\begin{align}
  &\ln f(\mathbf{h}_{b,\mathrm{MU}}[k]) = \mathbb{E}_{\bar{\gamma}_k, \zeta}\big\{\ln g(\mathbf{y}_{\mathrm{MU}_\mathcal{P}}[k]|\mathbf{h}_{b,\mathrm{MU}}[k],\zeta) + \ln g(\mathbf{h}_{b,\mathrm{MU}}[k]|\bar{\gamma}_k) + \ln g(\bar{\gamma}_k) + \ln g(\zeta) \big\} \label{solveh},\\
 \mathrm{ln} \; f(\mathbf{h}_{b,\mathrm{MU}}[k])& = \mathbb{E}_{\bar{\gamma}_k,\zeta} \Big\{-\zeta  (\mathbf{y}_{\mathrm{MU}_\mathcal{P}}[k]-\boldsymbol{\Lambda}_{\mathrm{MU}_\mathcal{P}}[k]\mathbf{h}_{b,\mathrm{MU}}[k])^H \mathbf{C}_w^{-1}(\mathbf{y}_{\mathrm{MU}_\mathcal{P}}[k]-\boldsymbol{\Lambda}_{\mathrm{MU}_\mathcal{P}}[k]\mathbf{h}_{b,\mathrm{MU}}[k]) - \frac{1}{2}\mathrm{Tr}\big[(\mathbf{\Gamma}_{k,\mathrm{MU}})\mathbf{h}_{b,\mathrm{MU}}[k]\mathbf{h}_{b,\mathrm{MU}}^T[k]\big] \Big\}. \label{lnh}
    \end{align}
    \hrulefill \vspace{-1.5\baselineskip}
\end{figure*}
\begin{figure*}[hbt!]
\begin{align}
    \mathrm{ln} \; f(\bar{\gamma}_k)& = \mathbb{E}_{\mathbf{h}_{b,\mathrm{MU}}[k]}\big\{\mathrm{ln}\;g(\mathbf{h}_{b,\mathrm{MU}}[k]|\bar{\gamma}_k) + \mathrm{ln} \; g(\bar{\gamma}_k) \big\} + \mathrm{const} \notag \\
    & = \big(\kappa-\frac{1}{2}\big) \sum_{\imath=1}^{U G_R G_{T,u}} \mathrm{ln}(\gamma_{k,i}) - \sum_{\imath = 1}^{U G_R G_{T,u}} (\chi + \mathbb{E}(|\mathbf{h}_{b,\mathrm{MU}}[k](\imath)|^2)\gamma_{k,\imath}) + \mathrm{const}. \label{post_hype}
\end{align}
\hrulefill \vspace{-1.5\baselineskip}
\end{figure*}
\begin{figure*}[hbt!]
    \begin{align}
        \mathrm{ln} \; f(\zeta)  & = \mathbb{E}_{\mathbf{h}_{b,\mathrm{MU}}[k]}\big\{\mathrm{ln} \; g(\mathbf{y}_{\mathrm{MU}_\mathcal{P}}[k]|\mathbf{h}_{b,\mathrm{MU}}[k],\zeta) + \mathrm{ln} \; g(\zeta) \big\} + \mathrm{const}, \label{postzeta}\\
        & = \mathbb{E}_{\mathbf{h}_{b,\mathrm{MU}}[k]}\big\{(\varphi - 1 + U G_R G_{T,u}) \mathrm{ln}(\zeta) + \zeta(\mathbf{y}_{\mathrm{MU}_\mathcal{P}}[k] - \boldsymbol{\Lambda}_{\mathrm{MU}_\mathcal{P}}[k]\mathbf{h}_{b,\mathrm{MU}}[k])^H \mathbf{C}_w^{-1} (\mathbf{y}_{\mathrm{MU}_\mathcal{P}}[k] - \boldsymbol{\Lambda}_{\mathrm{MU}_\mathcal{P}}[k]\mathbf{h}_{b,\mathrm{MU}}[k]) + \varsigma \big\} + \mathrm{const} \label{zetah}.
    \end{align}
    \hrulefill \vspace{-1.2\baselineskip}
\end{figure*}
\hspace{-3mm} However, such sophisticated probabilistic model does not admit exact Bayesian inference, as the posterior distribution of the unknown variables cannot be obtained in close-form. Consequently, we approximate the Bayesian inference by applying the \textit{mean-field} approximation as outlined in Eq. \eqref{VBLapprox}, with their individual posterior updates detailed below.\\
\hspace{-4.5mm} \underline{\textbf{Computing the posterior of} $\mathbf{h}_{b,\mathrm{MU}}[k]$:} Taking the logarithm on both sides of Eq. \eqref{solveh} and solving further yields Eq. \eqref{lnh}. Therefore, the resulting function is quadratic and can be approximated by a standard Gaussian, expressed as $f(\mathbf{h}_{b,\mathrm{MU}}[k]) \approx \mathcal{CN}(\boldsymbol{\mu}_k,\mathbf{\Sigma}_k)$ where the \textit{a posterior} mean and covariance are
\vspace{-2mm}
\begin{align}
    \boldsymbol{\Sigma}_k &= \big(\mathbb{E}(\mathbf{\Gamma}_{k,\mathrm{MU}})+\mathbb{E}(\zeta)\boldsymbol{\Lambda}_{\mathrm{MU}_{\mathcal{P}}}^H[k]\mathbf{C}_w^{-1}\boldsymbol{\Lambda}_{\mathrm{MU}_{\mathcal{P}}}^H[k]\big)^{-1}, \notag \\
    \boldsymbol{\mu}_k &= \mathbb{E}(\zeta)\boldsymbol{\Sigma}_k\boldsymbol{\Lambda}_{\mathrm{MU}_{\mathcal{P}}}^H[k]\mathbf{C}_w^{-1} \mathbf{y}_{\mathrm{MU}_{\mathcal{P}}}[k]. \label{E-step}
\end{align}
\underline{\textbf{Computing the posterior of} $\bar{\gamma}_k$}: By isolating the terms dependent on $\bar{\gamma}_k$ and taking the logarithm on both sides, we derive Eq. \eqref{post_hype}. Comparing with the Gamma distribution parameters, we obtain
\vspace{-2.5mm}
\begin{align}
    \tilde{\kappa} = \kappa; \;
    \tilde{\chi} = \chi + \mathbb{E}\{|\mathbf{h}_{b,\mathrm{MU}}[k](\imath)|^2\}, \label{M1}
\end{align}
where $\mathbb{E}\big\{|\mathbf{h}_{b,\mathrm{MU}}[k](\imath)|^2\big\} = |\boldsymbol{\mu}_k(\imath)|^2 + \mathbf{\Sigma}_k(\imath,\imath)$.\\
\underline{\textbf{Computing the posterior of} $f(\zeta)$:} Similarly, by rearranging the terms involving $\zeta$ and applying the logarithm to both sides, we establish Eq. \eqref{postzeta}, which consequently leads to Eq. \eqref{zetah}. Therefore, by aligning the resulting expression with the standard Gamma distribution parameters, we obtain
\vspace{-2mm}
\begin{align}
    \tilde{\varphi} &= (\varphi-1 + UG_RG_{T,u}), \notag\\
    \tilde{\varsigma} &= \varsigma + \mathbb{E}_{\mathbf{h}_{b,\mathrm{MU}}[k]}\big\{\big(\mathbf{y}_{\mathrm{MU}_\mathcal{P}}[k]-\boldsymbol{\Lambda}_{\mathrm{MU}_{\mathcal{P}}}[k]\mathbf{h}_{b,\mathrm{MU}}[k]\big)^H \notag \\ &\mathbf{C}_w^{-1} \big(\mathbf{y}_{\mathrm{MU}_\mathcal{P}}[k]-\boldsymbol{\Lambda}_{\mathrm{MU}_{\mathcal{P}}}[k]\mathbf{h}_{b,\mathrm{MU}}[k]\big)\big\}. \label{M2}
\end{align}
Note that the resulting variational inference algorithm not only provides closed-form updates for each variable, but also accounts for the uncertainties associated with the estimated parameters.\\
\underline{\textbf{Hyperparameter setting:}} We assume no prior knowledge where the hyperparameters $\{\kappa,\chi\}_{t=1}^{UG_RG_{T,u}}$ and $\{\varphi,\varsigma\}$ are set as $10^{-6}$ to represent non-informative priors. 
A similar procedure can be applied to the USP-DCE-based approach for sparse estimation; however, we have omitted it here due to space constraints. Algorithm-\ref{est_algo} provides a summary of the sparse estimation procedure. The final estimate is given by
\vspace{-2mm}
\begin{align}
    \widehat{\mathbf{H}}_{\mathrm{VBI}} = \mathbf{A}_R(\Theta_R, f_k)\mathrm{vec}^{-1}(\widehat{\mathbf{h}}_{\mathrm{VBI}}[k])\mathbf{A}_{T,\mathrm{MU}}^H(\Phi_T,f_k), \notag
\end{align}
where the concatenated transmit vector $\mathbf{A}_{T,\mathrm{MU}}^H(\Phi_T,f_k) = \mathrm{blkdiag}(\mathbf{A}_{T,1}(\Phi_T,f_k),\cdots,(\mathbf{A}_{T,U}(\Phi_{T,U},f_k)) \in \mathbb{C}^{N_T G_T \times N_T G_T}$.
\vspace{-5mm}
\section{Simulation Results} \label{simu_results}
\vspace{-1mm}
We consider a practical scenario of $U = 3$ users each employing $N_{T,u} = 4$ transmit antennas possessing $N_{\mathrm{RF},u} = 2$ RF chains. The BS is equipped with $N_R = 64$ antennas with $N_{\mathrm{RF}}^R = 8$ RF chains, where each subarray possess $N_R^{l} = \frac{N_R}{N_{\mathrm{RF}}^R} = 8$ antennas. The carrier frequency is $f_c = 0.671 \, \mathrm{THz}$ and transmission distance is $d = 12 \, \mathrm{m}$ with molecular absorption $k_\mathrm{abs}(f_k) = 0.01358 \, \mathrm{m}^{-1}$. We consider a single LoS path and $3$ NLoS path with $N_{\mathrm{ray}} = 1$ diffused ray. Furthermore, we consider $K = 64$ subcarriers with $B = 10$ GHz and $M = 15$ pilot blocks. The transmit dictionary includes $G_{T,u} = 8 \, \mathrm{atoms}$ while each subarray includes $G_R^{l} = 16 \, \mathrm{atoms}$. The transmit and receive antenna gains are $\sum_{u=1}^U \mathfrak{B}_{T,u} = 16 \, \mathrm{dBi}$ and $\mathfrak{B}_R^{l} = 16 \, \mathrm{dBi}$. The frequency-independent phase shifters are designed as described in \cite{garg2025bayesian}. Furthermore, we use Gaussian mixture model (GMM) to generate the AoAs/AoDs due to its accurate representation in the THz band \cite{garg2025bayesian}.
\begin{algorithm}[t]
\DontPrintSemicolon 
\KwIn{$\mathbf{y}_{\mathrm{MU}_{\mathcal{P}}}[k], \boldsymbol{\Lambda}_{\mathrm{MU}_{\mathcal{P}}}[k],\mathbf{C}_w^{-1}, \epsilon, N_{\mathrm{max}}$}
\textbf{Initialization:} $\gamma_{k,\imath}^{(0)} = 1 \; \forall \; 0 \leq \imath \leq U G_R G_{T,u}; \widehat{\mathbf{\Gamma}}_{k,\mathrm{MU}}^{(0)} = \mathbf{I}_{U G_R G_{T,u}}; \widehat{\mathbf{\Gamma}}_{k,\mathrm{MU}}^{(-1)} = \mathbf{0}; j = 0$
 
\While{$\parallel \widehat{\mathbf{\Gamma}}_{k,\mathrm{MU}}^{(j)} - \widehat{\mathbf{\Gamma}}_{k,\mathrm{MU}}^{(j-1)} \parallel_{\mathcal{F}}^2 \geq \epsilon$ and $j < N_{\mathrm{max}}$}
{
$j = j+1$

Update the \textit{a posteriori} mean and covariance using Eq. \eqref{E-step}

Update the hyperparameters and noise covariance using Eq. \eqref{M1} and \eqref{M2} respectively.}

\textbf{Output:~~}{$\widehat{\mathbf{h}}_{\mathrm{VBI}}[k] = \boldsymbol{\mu}_k$}
\caption{VBI based sparse channel estimation}
\label{est_algo}
\end{algorithm}
The NMSE metric is defined as $\mathrm{NMSE} = \frac{\sum_{k=0}^{K-1} \parallel \widehat{\mathbf{H}}[k]-\mathbf{H}[k]\parallel^2_F}{\sum_{k=0}^{K-1}\parallel \mathbf{H}[k] \parallel^2_F}$.
\begin{figure*}
	\centering
	\subfloat[]{\includegraphics[scale=0.32]{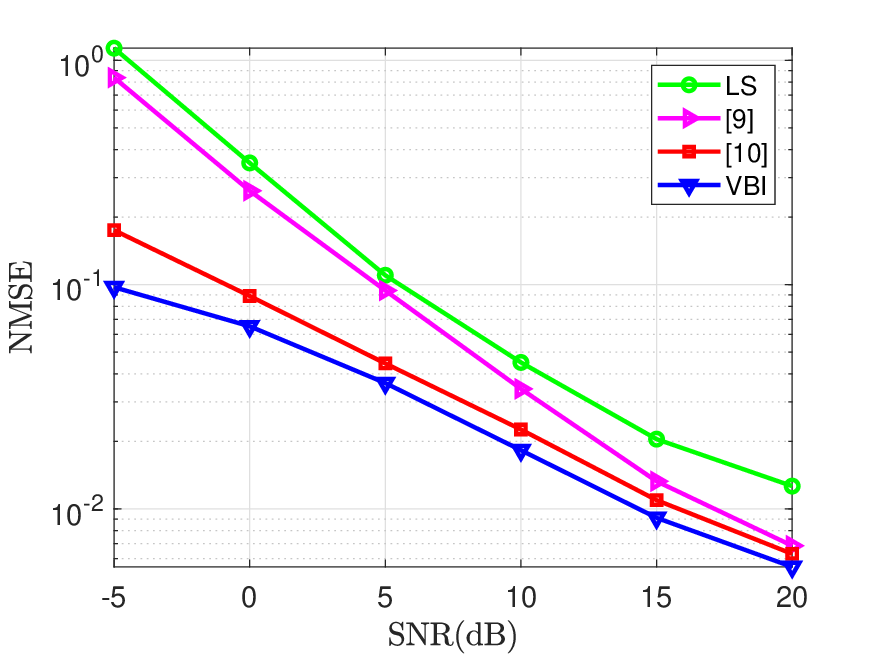}}
	\hfil
	\hspace{-12pt}\subfloat[]{\includegraphics[scale=0.32]{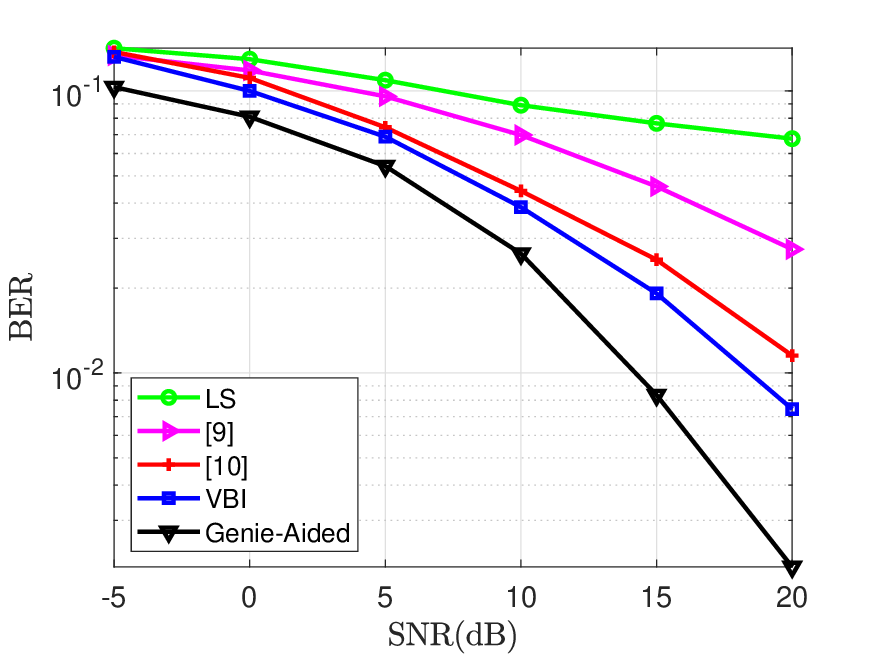}}
 	\hfil
	\hspace{-12pt} \subfloat[]{\includegraphics[scale=0.32]{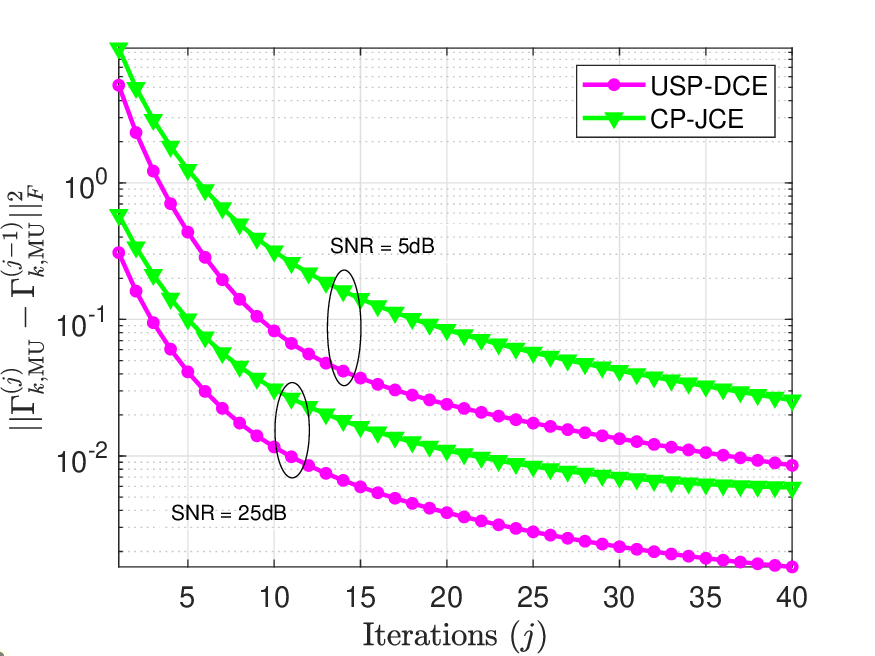}}
	\hfil
	\hspace{-12pt}\subfloat[]{\includegraphics[scale=0.32]{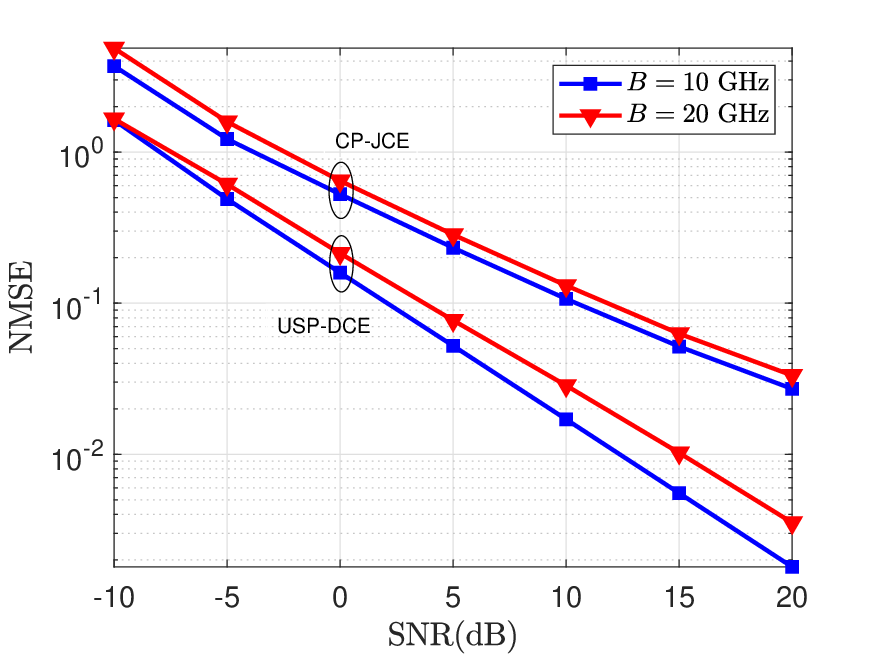}}
    \vspace{-2mm}
	\caption{Performance comparison between the proposed approach and existing state-of-the-art techniques $\left(a\right)$ NMSE vs SNR $ \left(b\right) $ BER vs SNR $ \left(c\right) $ $\parallel \boldsymbol{\Gamma}^{(j)}_{k,\mathrm{MU}} - \boldsymbol{\Gamma}^{(j-1)}_{k,\mathrm{MU}} \parallel_\mathcal{F}^2$ vs iterations for the proposed CP-JCE and USP-DCE system model $ \left(d\right) $ NMSE vs SNR for the proposed CP-JCE and USP-DCE system model by varying bandwidth $B \in \{10,20\} \:$GHz.\vspace{-1.8 \baselineskip}}
	\label{CP-USP} 
\end{figure*} 
Fig. \ref{CP-USP}(a) illustrates the NMSE performance as a function of SNR. The poor performance of the least squares (LS) method is attributed to its inability to incorporate sparsity priors into the estimation process, which poses a significant limitation compared to CS-based approaches. Moreover, the algorithm in \cite{aghda2023superimposed} is prone to both structural and convergence errors. In particular, setting a low stopping threshold introduces structural errors, whereas a high stopping threshold results in convergence errors. Additionally, the algorithm in \cite{talaei2018hybrid} frequently converges to suboptimal local minima, leading to convergence errors. In contrast, variational Bayes inference assumes parameterized prior distributions and optimizes the variational parameters to approximate the posterior distribution, thereby enabling the joint estimation of all unknown variables and making it well-suited for diverse propagation environments. Fig. \ref{CP-USP}(b) depicts the accuracy of data detection achieved by employing the MMSE receiver with the proposed Bayesian inference, along with other conventional sparse estimation based techniques and the Genie-aided detector as benchmarks. As illustrated, the resulting BER decreases consistently with increasing SNR, which closely aligns with the NMSE results presented previously, underscoring the high quality of the CSI estimated by the Bayesian inference. Additionally, the BER of the proposed channel learning approach closely matches the hypothetical Genie-aided detector (derive from true channel knowledge) at higher SNRs, highlighting its efficacy and robustness in realistic scenarios. Fig. \ref{CP-USP}(c) depicts the convergence of the variational Bayes inference as a function of iterations for the CP-JCE and USP-DCE based models. To achieve a target NMSE of $0.1$, the USP-DCE model requires $9$ iterations, while the CP-JCE model requires $19$ iterations at $\mathrm{SNR} = 5 \: \mathrm{dB}$. Note that, the complexity increase in CP-JCE remains moderate. This is because CP-JCE effectively exploits the common angular support shared among users, reducing redundancy and preventing complexity from scaling linearly with user count.
\vspace{-3mm}
\subsection{Trade-off between the CP-JCE and USP-DCE approaches}
\subsubsection{Complexity} The CP-JCE approach utilizes a large channel matrix that includes the channels of all users, resulting in a higher computational complexity. In contrast, the USP-DCE method employs smaller channel matrices specific to individual users, leading to reduced complexity. Specifically, the complexity of CP-JCE is of the order $\mathcal{O}(U^3 G_{T,u}^3 G_R^3)$, while the complexity of USP-DCE is of the order of $\mathcal{O}(U G_{T,u}^3 G_R^3)$. This demonstrates that USP-DCE has a complexity advantage over CP-JCE, particularly as the number of users increases.
\vspace{1mm}
\subsubsection{NMSE performance} From Fig. \ref{CP-USP}(d), it can be observed that the NMSE performance of the CP-JCE approach is poorer compared to the USP-DCE approach. This is attributed to the fact that the CP-JCE approach handles $U$ users simultaneously, whereas the USP-DCE method is designed for a single user. Consequently, the increased number of users in CP-JCE leads to higher estimation errors due to the added complexity and interference in the estimation process.
\subsubsection{Coherence-Time} In the CP-JCE approach, only two precoding matrices are required i.e., one for pilot decoupling and other for data decoupling for all the users. In contrast, the USP-DCE approach requires $2 \times U$ precoding matrices i.e., $U$ matrices for pilot decoupling and $U$ matrices for data decoupling. As a result, the CP-JCE approach benefits from a \textit{shorter required coherence time}, making it more suitable for rapidly changing channels, whereas the longer coherence time of the USP-DCE approach can be a disadvantage in dynamic environments.
\vspace{-2mm}
\section{Conclusion}
\vspace{-1mm}
This study proposes affine precoded SIP models based on CP-JCE and USP-DCE for partially connected MU THz systems and analyzes the trade-offs between them in terms of complexity and estimation accuracy. A variational Bayes inference framework is developed for efficient estimation by considering the dual-wideband channel. Simulation results demonstrate the performance gains achieved by the channel estimation approach.
\vspace{-0.7 \baselineskip}
\bibliographystyle{IEEEtran}
\bibliography{References}
\end{document}